\documentclass[12pt,a4paper]{article}

\usepackage{epsfig}

\begin{document}


\noindent
DESY 04-100, SFB/CPP-04-17, CPT-2004-P-038
\hfill{\tt hep-lat/0406027\_v2}\\
June 2004
\vspace{20pt}

\begin{center}
{\LARGE\bf Staggered eigenvalue mimicry}
\end{center}
\vspace{10pt}

\begin{center}
{\large
{\bf Stephan D\"urr}$\,{}^{a}$\hspace{2pt},\hspace{6pt}
{\bf Christian Hoelbling}$\,{}^{b}$\hspace{8pt}and\hspace{6pt}
{\bf Urs Wenger}$\,{}^{c}$
}
\\[10pt]
${}^a\,$DESY Zeuthen, Platanenallee 6, D-15738 Zeuthen, Germany\\
${}^b\,$Centre de Physique Th\'eorique, Case 907, CNRS Luminy,
F-13288 Marseille Cedex 9, France\\
${}^c\,$NIC/DESY Zeuthen, Platanenallee 6, D-15738 Zeuthen, Germany\\
\end{center}
\vspace{10pt}

\begin{abstract}
\noindent
We study the infrared part of the spectrum for UV-filtered staggered Dirac
operators and compare them to the overlap counterpart. With sufficient
filtering and at small enough lattice spacing the staggered spectra manage to
``mimic'' the overlap version. They show a 4-fold near-degeneracy, and a clear
separation between would-be zero modes and non-zero modes.
This suggests an approximate index theorem for filtered staggered fermions
and a correct sensitivity to the topology of QCD. Moreover, it supports
square-rooting the staggered determinant to obtain dynamical ensembles with
$N_f\!=\!2$.
\end{abstract}


\hyphenation{topo-lo-gi-cal simu-la-tion theo-re-ti-cal mini-mum}


\newcommand{\pad}{\partial}
\newcommand{\pas}{\partial\!\!\!/}
\newcommand{\Dsl}{D\!\!\!\!/\,}
\newcommand{\Psl}{P\!\!\!\!/\;\!}
\newcommand{\hqu}{\hbar}
\newcommand{\ovr}{\over}
\newcommand{\til}{\tilde}
\newcommand{\pri}{^\prime}
\renewcommand{\dag}{^\dagger}
\newcommand{\<}{\langle}
\renewcommand{\>}{\rangle}
\newcommand{\gaf}{\gamma_5}
\newcommand{\lap}{\triangle}
\newcommand{\trc}{{\rm tr}}

\newcommand{\al}{\alpha}
\newcommand{\be}{\beta}
\newcommand{\ga}{\gamma}
\newcommand{\de}{\delta}
\newcommand{\ep}{\epsilon}
\newcommand{\ve}{\varepsilon}
\newcommand{\ze}{\zeta}
\newcommand{\et}{\eta}
\renewcommand{\th}{\theta}
\newcommand{\vt}{\vartheta}
\newcommand{\io}{\iota}
\newcommand{\ka}{\kappa}
\newcommand{\la}{\lambda}
\newcommand{\rh}{\rho}
\newcommand{\vr}{\varrho}
\newcommand{\si}{\sigma}
\newcommand{\ta}{\tau}
\newcommand{\ph}{\phi}
\newcommand{\vp}{\varphi}
\newcommand{\ch}{\chi}
\newcommand{\ps}{\psi}
\newcommand{\om}{\omega}

\newcommand{\psb}{\overline{\psi}}
\newcommand{\etb}{\overline{\eta}}
\newcommand{\psd}{\psi^{\dagger}}
\newcommand{\etd}{\eta^{\dagger}}
\newcommand{\beq}{\begin{equation}}

\newcommand{\eeq}{\end{equation}}
\newcommand{\bdm}{\begin{displaymath}}
\newcommand{\edm}{\end{displaymath}}
\newcommand{\bea}{\begin{eqnarray}}
\newcommand{\eea}{\end{eqnarray}}

\newcommand{\mr}{\mathrm}
\newcommand{\mb}{\mathbf}
\newcommand{\Nf}{{N_{\!f}}}
\newcommand{\Nc}{{N_{\!c}}}
\newcommand{\ri}{\mr{i}}
\newcommand{\DW}{D_\mr{W}}
\newcommand{\DN}{D_\mr{N}}

\newcommand{\MeV}{\,\mr{MeV}}
\newcommand{\GeV}{\,\mr{GeV}}
\newcommand{\fm}{\,\mr{fm}}


\section{Introduction}


Recent years have brought a remarkable increase of interest in dynamical
($\Nf\!=\!2$ or $\Nf\!=\!2\!+\!1$) QCD runs employing staggered sea-quarks.
One of the conceptual issues with these efforts is that the staggered quark
action is not doubler-free; a single staggered field will eventually generate
four flavors in the continuum.
Hence, to create an ensemble with $\Nf\!=\!2$, one simply uses the square-root
of the staggered determinant, and for $\Nf\!=\!2\!+\!1$ an additional
quartic-root with a higher mass.
The problem is that taking a fractional power of the determinant is, in
general, not a legitimate operation in quantum field-theory, contrary to
positive integer powers.
The worst-case scenario is, therefore, that these runs might not represent an
ab-initio approach to QCD in the strong coupling regime.
In this respect it does not help that the recent staggered $\Nf\!=\!2\!+\!1$
results with light sea-quarks look very promising~\cite{Davies:2003ik}.
It has been clearly pointed out by Jansen at the Tsukuba
conference~\cite{Jansen:2003nt} and in DeGrand's review~\cite{DeGrand:2003xu}
that --~if the square-rooting trick cannot be justified~-- dynamical staggered
runs with $\Nf\!\notin\!4\mr{I\!N}$ must be considered a \emph{model\/} of QCD.
The results in~\cite{Davies:2003ik} show that it would be a much better model
than quenched QCD, but from a fundamental point of view little is won if the
answer would be that there is no justification for the fractional power.

\bigskip

The goal of this note is to confront the low-lying eigenvalues of the massless
staggered Dirac operator~\cite{staggered} with analogous spectra of a
``benchmark'' operator which is known to be without conceptual problems.
We choose the massless overlap operator~\cite{overlap}, since it satisfies
(like domain-wall~\cite{domainwall} and classically perfect~\cite{perfect}
fermions) the Ginsparg-Wilson relation
\cite{Ginsparg:1981bj}
\beq
D\gaf+\gaf D={1\ovr\rh}D\gaf D
\label{ginspargwilson}
\eeq
with $\rho$ a parameter to be specified later.
This relation is of prime interest, because it implies invariance of the action
(at finite lattice spacing) under the adapted chiral
transformation~\cite{Luscher:1998pq}
\beq
\de\ps=\gaf(1-{1\ovr\rh}D)\ps,\qquad\de\psb=\psb\gaf
\label{luscher}
\;,
\eeq
which in turn excludes additive mass renormalization and prevents operators in
different chiral multiplets from mixing.
On the other hand, the staggered action breaks the full $SU(\Nf\!=\!4)_A$
flavor symmetry, but the remnant $U(1)$-symmetry
\beq
\ch(x)\to\exp\Big(\ri\,\th_A(-1)^{\sum x_\nu}\Big)\ch(x)
\;,\quad
\bar\ch(x)\to\bar\ch(x)\exp\Big(-\ri\,\th_A(-1)^{\sum x_\nu}\Big)
\;,
\label{stagremnant}
\eeq
still protects the fermion mass against additive renormalization.

A point worth emphasizing is that modern staggered simulations use highly
``improved'' staggered quarks.
These are operators where one replaces in the usual definition
\begin{equation}
D^\mr{stag}=
{1\ovr2}\sum_{\mu}
\et_\mu(x)
\Big(
U_\mu(x)\de_{x+\hat\mu,y}-U_\mu\dag(x\!-\!\hat\mu)\de_{x-\hat\mu,y}
\Big)
\end{equation}
[with $\et_\mu(x)\!=\!(-1)^{\sum_{\nu<\mu}x_\nu}$]
the parallel transporter $U_\mu(x)$ by a weighted sum of several
gauge-covariant paths from $x$ to $x\!+\!\hat\mu$, and there is an abundance of
proposals which terms should enter and what are useful
weights~\cite{uvfilteredstag}.
From the Symanzik point of view~\cite{Symanzik}, these actions are in general
\emph{not~improved\/}~-- they are in the same class with $O(a^2)$ cut-off
effects as the original staggered action (albeit the extrapolation slope is
typically reduced, and the scaling window might begin earlier).
Since these modifications are designed to reduce the dramatic background
fluctuations at the cut-off level (in particular the ``taste-changing''
interactions due to highly virtual gluon exchanges~\cite{uvfilteredstag}), one
should rather speak of ``UV-filtered'' staggered quarks.

The massless overlap operator is defined through~\cite{overlap}
\beq
D^\mr{over}=
\rh\Big(
1+D^\mr{W}_{-\rho}({D^\mr{W}_{-\rho}}^{\!\dagger} D^\mr{W}_{-\rho})^{-1/2}
\Big)
=\rh
\Big(
1+\gaf\,\mr{sign}(\gaf D^\mr{W}_{-\rho})
\Big)
\label{diracoverlap}
\eeq
with $D^\mr{W}_{-\rho}$ the Wilson operator at negative mass $-\rho$.
In the free field limit $\rho\!=\!1$ is appropriate, but for finite $\be$ one
usually chooses some $\rho\!>\!1$ to optimize the locality of
$D^\mr{over}$~\cite{Hernandez:1998et}.
Given the success of the UV-filtered staggered operators, it is natural to
consider modified overlap operators where the Wilson Kernel is constructed from
smoothed parallel-transporters, too
\cite{DeGrand:2000tf,Durr:2003xs}.
The resulting filtered overlap operator is still in the Symanzik class with
$O(a^2)$ artefacts, if one stays at a fixed smearing level for all $\be$.
In other words: The choice of the covariant derivative in $D^\mr{W}$ is an
$O(a^2)$ ambiguity in $D^\mr{over}$ like the choice of $\rh$.
We thus take the liberty to fix $\rh\!=\!1$ at all $\be$-values (except for one
check), and to just change the smearing level.

\bigskip

With this setup we are ready for an investigation, which mainly follows the
``spectral~hint'' section of Ref.~\cite{Durr:2003xs}.
To come back to the starting point, what one would like to check is whether
UV-filtered staggered quarks develop approximately 4-fold degenerate
eigenvalues which coincide with the overlap counterparts such that the
rooted staggered determinant generates, up to $O(a^2)$ cut-off effects, the
correct ensemble with 1 or 2 light sea-quarks,
\begin{equation}
\det D^\mr{over}\propto
(\det D^\mr{stag})^{1/4}+O(a^2)
\;.
\label{fundamental1}
\end{equation}
Our approach is to focus on the technically simpler version
\begin{equation}
\det D^\mr{over}_\mr{UV-filtered}\propto
(\det D^\mr{stag}_\mr{UV-filtered})^{1/4}+O(a^2)
\;,
\label{fundamental2}
\end{equation}
which is equivalent, since the two left- and the right-hand sides differ by a
trivial rescaling and $O(a^2)$ terms only.
If such a relationship can be shown, the overlap operator would be a local
operator with the same determinant, up to cut-off effects, as the rooted
staggered operator, hence providing a bypass to the locality issue for the
valence quarks presented in~\cite{Bunk:2004br}.


\section{Low-lying eigenvalues of UV-filtered Dirac operators}


The plan is to compare --~configuration by configuration~-- the low-lying
eigenvalue spectrum of a UV-filtered staggered Dirac operator against that
of a filtered overlap operator.

Regarding the filtering, we decided not to follow the details of current
state-of-the-art staggered fermions, but to concentrate on two prototype
constructs, the APE~\cite{ape} and HYP~\cite{hyp} smoothed links with the
$SU(3)$-projection of Ref.~\cite{Kiskis:2003rd} which are used for both
types of lattice fermions.
Throughout, we used the parameters $\al_\mr{APE}\!=\!0.5$~\cite{ape} and 
$\al_\mr{HYP}\!=\!(0.75, 0.6, 0.3)$~\cite{hyp}.
By comparing the two recipes and the various iteration levels, we will be able
to judge how much our results are robust against changing the details of the
smearing procedure.

The key issue is whether these filtered staggered Dirac operators develop an
approximate 4-fold degeneracy, as it has been demonstrated in great detail in
2D (where the degeneracy is 2-fold)~\cite{Durr:2003xs},
and seen more recently in 4D~\cite{Follana:2004sz}.

\subsection{Dynamical lattices}

Let us begin with a full QCD set that was used for spectroscopy.
CU\_0001 is an ensemble of $49$ configurations with $16^3\times32$ geometry,
generated (via the square-root trick) with $2$ ``tastes'' of staggered quarks
at bare mass $m\!=\!0.01$ and coupling $\be\!=\!5.7 $.
The lattice spacing is around $a\!\simeq\!0.1\fm$.
We downloaded it from~\cite{nersc}, the pertinent original publication is
\cite{Brown:1991qw}.

\begin{table}[b]
\begin{center}
\begin{tabular}{|ccccc|}
\hline
u\_CU\_0001.0500 & conf\_000 & 0.577307 & $+0.9469$ &              $+1$\\
u\_CU\_0001.1200 & conf\_006 & 0.577942 & $-1.8857$ &              $-2$\\
u\_CU\_0001.2000 & conf\_014 & 0.577612 & $+0.0248$ & $(-1,-1,+1,0,+1)$\\
u\_CU\_0001.3200 & conf\_026 & 0.577187 & $-0.0981$ &               $0$\\
u\_CU\_0001.4400 & conf\_038 & 0.577232 & $+0.8983$ & $(+1,+1,+1,0,+1)$\\
u\_CU\_0001.5400 & conf\_048 & 0.577220 & $-4.4203$ &              $-5$\\
\hline
\end{tabular}
\end{center}
\vspace{-6mm}
\caption{\sl Original name, nickname, plaquette, $q_\mr{nai}$ and $n_-\!-\!n_+$
of the CU\_0001 configurations. Whenever they would not agree the indices for
the standard, 1\,APE, 3\,APE, 1\,HYP and 3\,HYP overlap operators are given
separately.}
\label{tab:CU0001}
\end{table}

In order to select configurations with and without overlap zero modes, we first
measured a lattice version of the continuum topological charge
$q_\mr{nai}={1\ovr32\pi^2}\sum_x\trc(\ep_{\mu\nu\si\rh}F_{\mu\nu}F_{\si\rh})$,
where $F_{\mu\nu}$ is defined via the clover-leaf construction involving
smeared links (7 HYP steps).
Based on this, we selected the configurations shown in Tab.~\ref{tab:CU0001}
for the spectral analysis.
The occurrence of configurations where the overlap index
$n_-\!-\!n_+\!=\!{1\ovr2\rh}\trc(\gaf D^\mr{over})$ (denoted $q$ in the
following) depends on the smearing level and type (or on the projection
parameter $\rh$) is of no concern -- it is a typical $O(a^2)$ effect.
The ambiguity is not removed by using vigorously filtered operators, but it
disappears in the continuum limit.

\begin{figure}
\vspace{-4mm}
\begin{center}
\epsfig{file=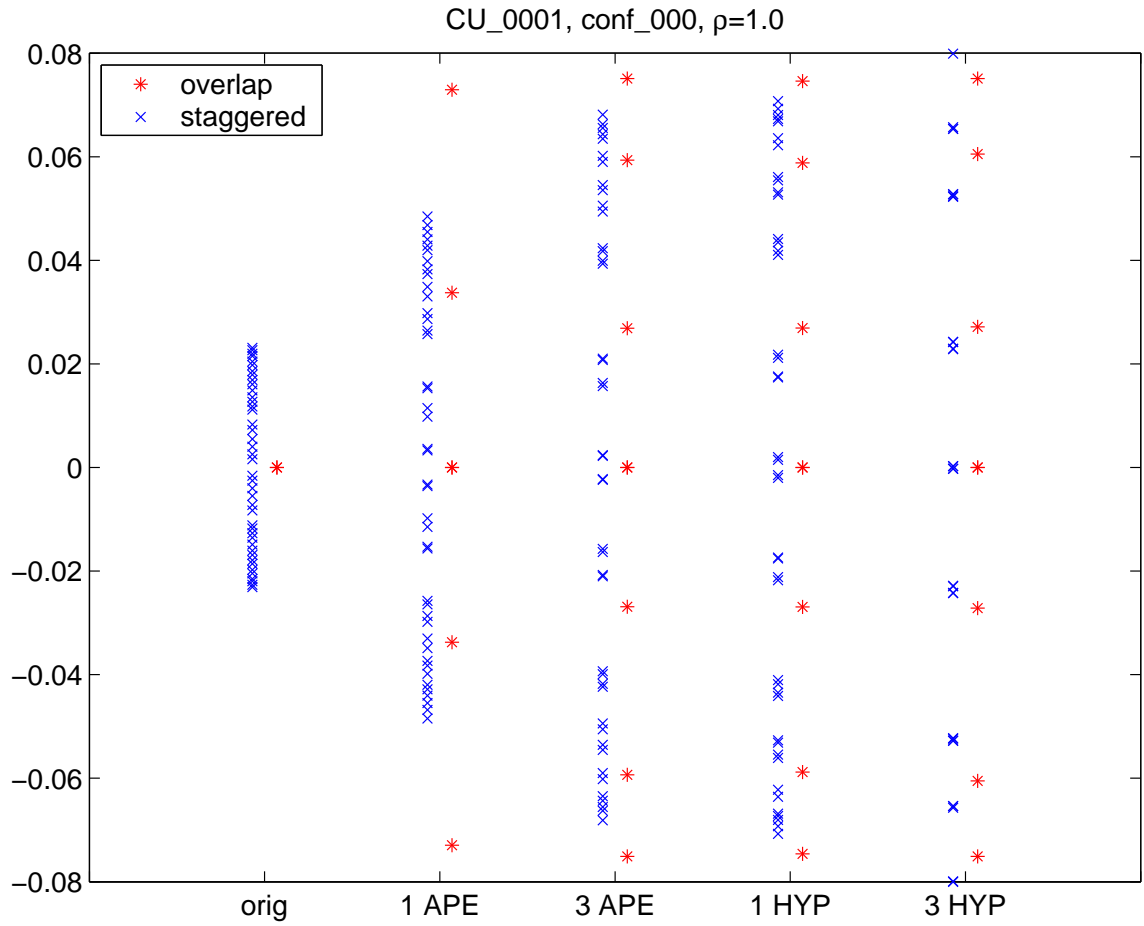,height=12.4cm}
\epsfig{file=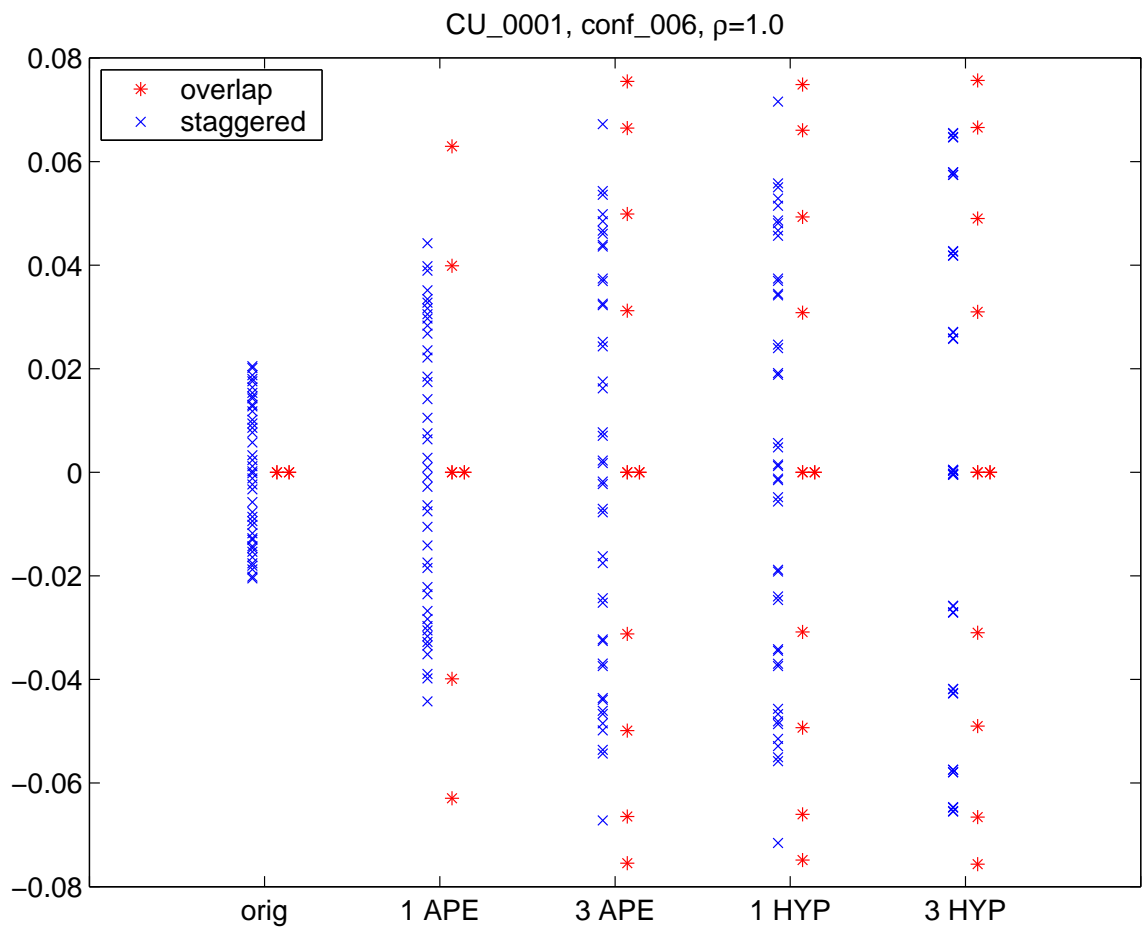,height=12.4cm}
\end{center}
\vspace{-8.4mm}
\caption{Eigenvalues on the configurations 00 and 06 in the ensemble CU\_0001.}
\label{fig:CU_uno}
\end{figure}

\begin{figure}
\vspace{-4mm}
\begin{center}
\epsfig{file=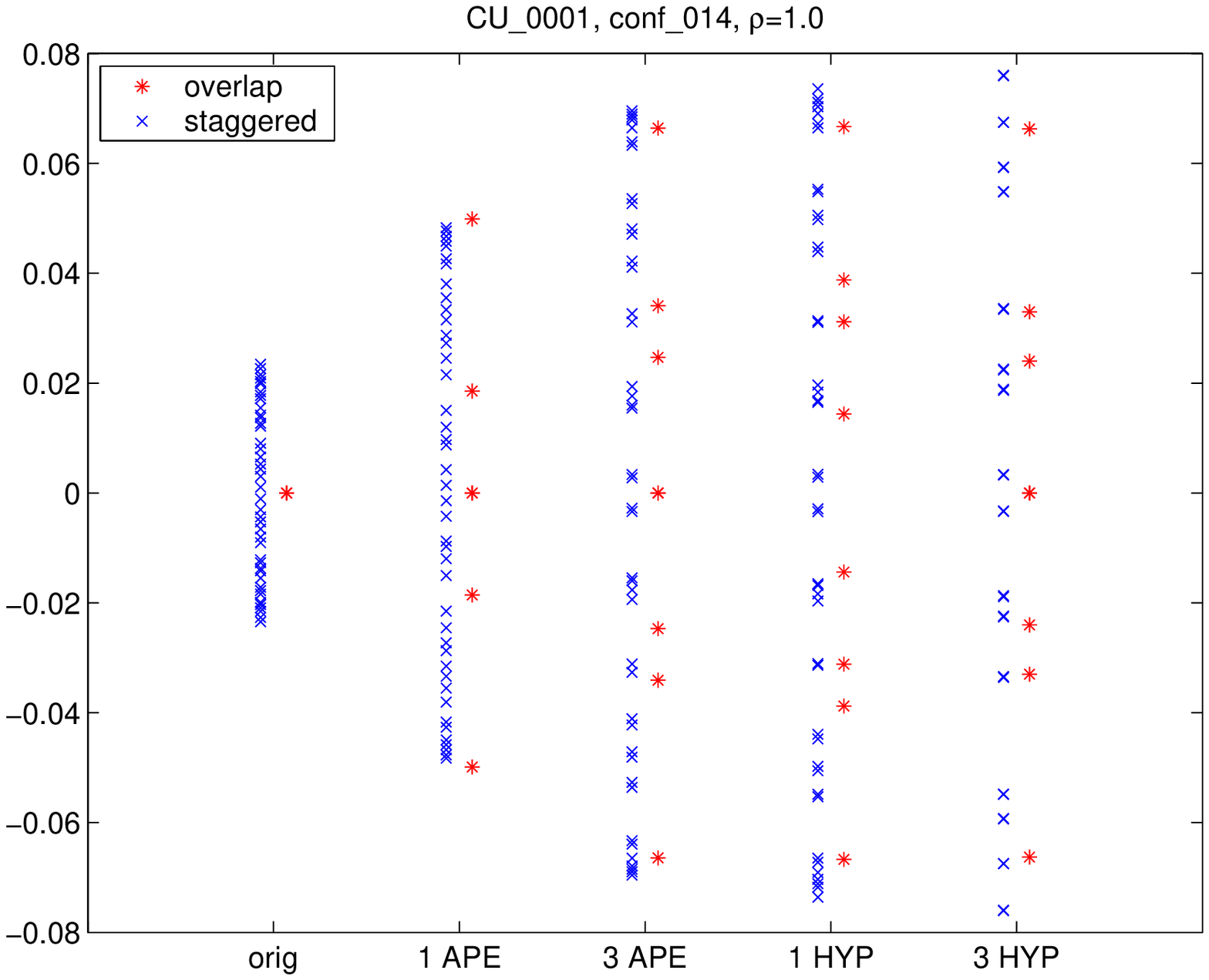,height=12.4cm}
\epsfig{file=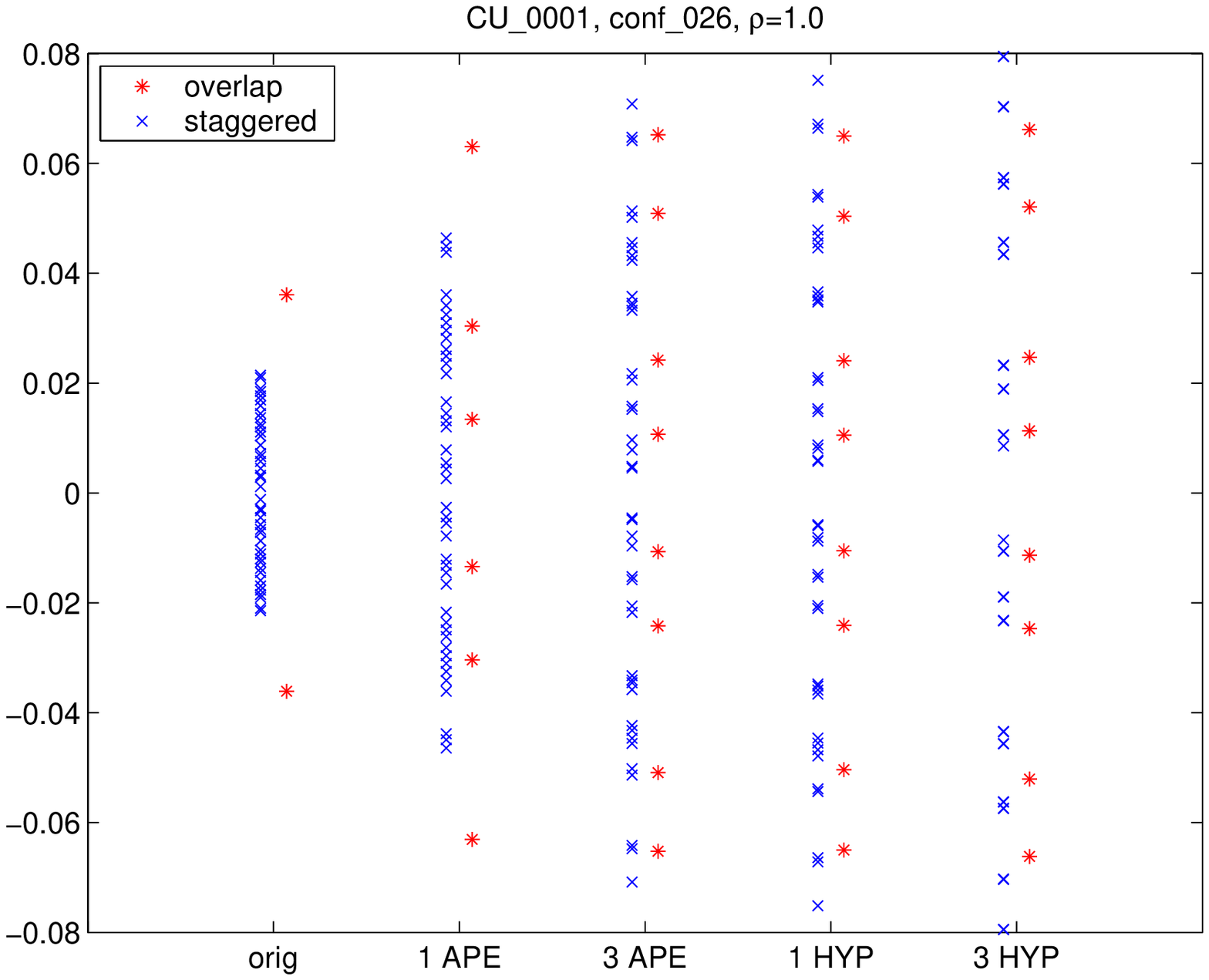,height=12.4cm}
\end{center}
\vspace{-8.4mm}
\caption{Eigenvalues on the configurations 14 and 26 in the ensemble CU\_0001.}
\label{fig:CU_due}
\end{figure}

\begin{figure}
\vspace{-4mm}
\begin{center}
\epsfig{file=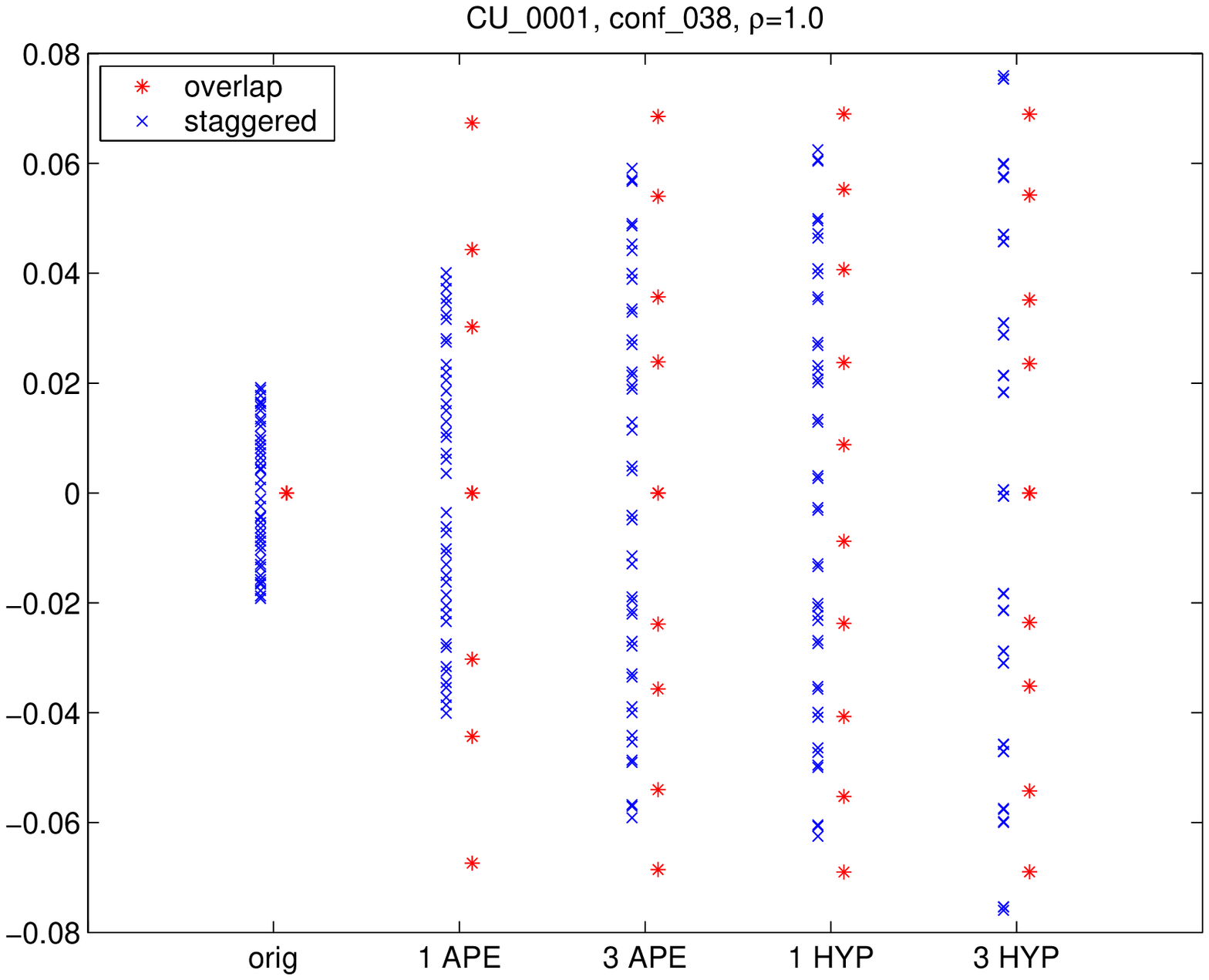,height=12.4cm}
\epsfig{file=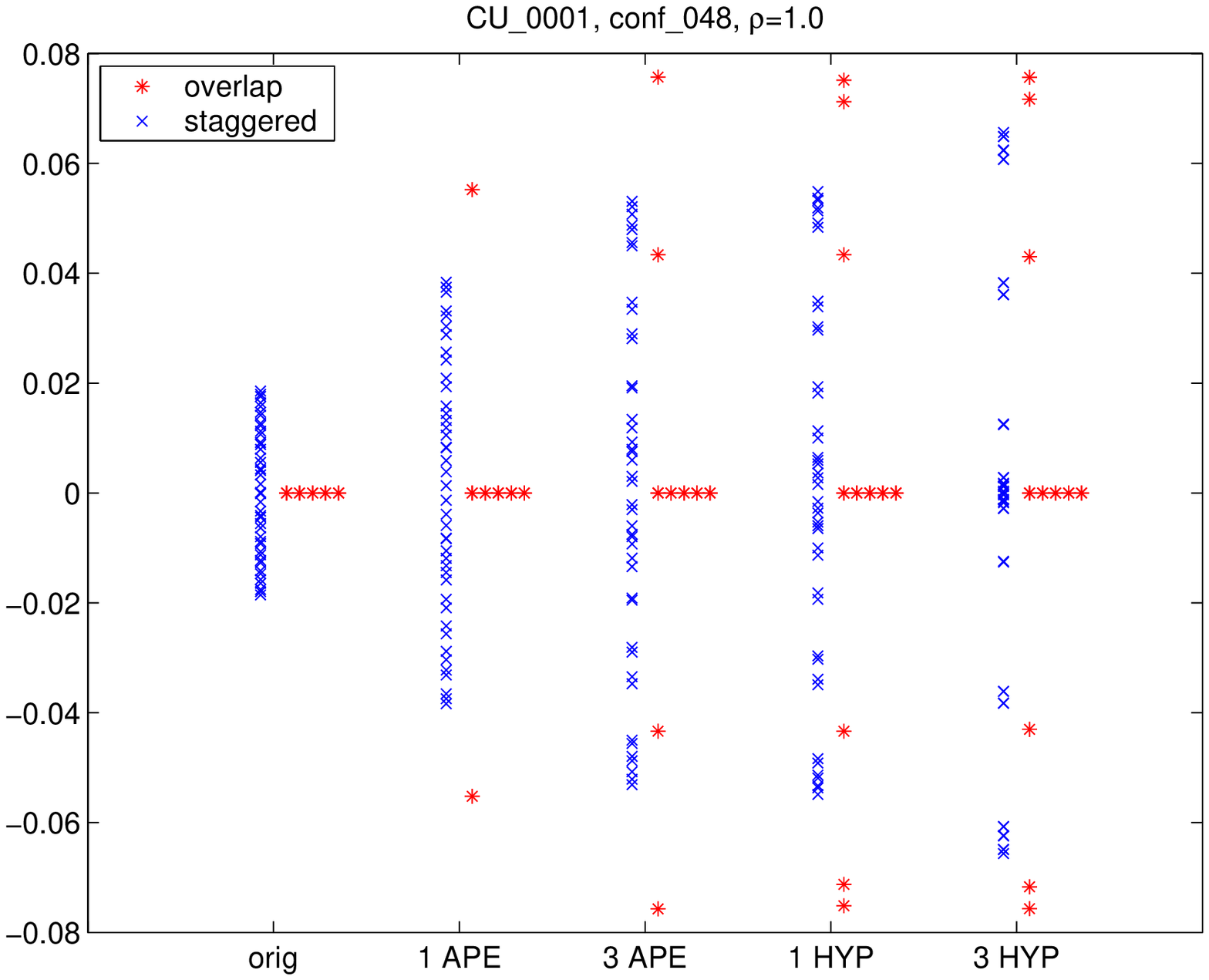,height=12.4cm}
\end{center}
\vspace{-8.4mm}
\caption{Eigenvalues on the configuration 38 and 48 in the ensemble CU\_0001.}
\label{fig:CU_tre}
\end{figure}

Since $D^\mr{stag}$ and $D^\mr{over}$ are normal, we determined the spectra of
the hermitean operators ${D^\mr{stag}}\dag D^\mr{stag}, {D^\mr{over}}\dag
D^\mr{over}$ and reconstructed the staggered and overlap eigenvalues,
respectively.
The eigenvalues of the hermitean operators have been determined with the
Ritz functional method~\cite{Kalkreuter:1995mm}.
We present the eigenvalues of $D^\mr{stag}$ and the chirally rotated
\begin{equation}
\hat\la=\Big(1/\la-1/(2\rh)\Big)^{-1}
\label{chiralrotated}
\end{equation}
of $D^\mr{over}$ which are both purely imaginary.
Figs.~\ref{fig:CU_uno},~\ref{fig:CU_due},~\ref{fig:CU_tre} contain our results.
The thin-link operators are compared on the left-most side, and the filtered
cousins with the same type of parallel-transporters are arranged next to each
other.
With $a^{-1}\!\simeq\!2\GeV$ the energy axis includes modes up to $160\MeV$.

The original operators show no similarity at all.
Upon employing one APE-smearing step, the staggered spectra get stretched
while the overlap-spectra get squeezed.
Still, there is no similarity, but the \emph{eigenvalue~densities\/} (divided
by $4$ in the staggered case) have been equalized quite a bit.
At intermediate filtering levels (3~APE and 1~HYP steps) all staggered
eigenvalues have grouped into almost-degenerate twin-pairs.
At the 3~HYP level, finally, a second grouping of twin-pairs into
\emph{quadruples of near-degenerate staggered eigenmodes\/} has taken place.
Stretching the staggered 3~HYP spectra with a factor $Z\!\simeq\!1.2$, one
finds a rather convincing agreement between the quadruples and the (individual)
overlap modes.
There is a clear \emph{gap\/} between staggered ``would-be'' zero modes and
non-zero modes, though this rule is challenged by some configurations.
On conf\_014 (and to some extent on conf\_038) the operator
$D^\mr{stag}_\mr{3 HYP}$ reproduces the confusion that the overlap has
regarding the charge of this configuration (cf.\ footnote~\ref{foot:thisone}).
And conf\_048 seems to be hard to resolve, simply for its high $|q|$.
Note, finally, the stability (on a clear majority of configurations) of the
low-lying overlap spectrum from 3~APE through 3~HYP.

All these findings are in complete analogy to what has been seen in great
detail in 2D~\cite{Durr:2003xs}.
The only difference is that the near-degeneracy is 2-fold there, since
(\ref{stagremnant}) is the lattice remnant of the $SU(\Nf\!=\!2)$ symmetry in
2D.


\subsection{Matched quenched lattices}

To check whether $O(a^2)$ effects really capture the difference between
filtered staggered and overlap spectra, we generated 4 sets of matched
quenched lattices, aiming at $L/r_0\!=\!2.24$ with the Sommer scale
$r_0\simeq0.5\fm$~\cite{Sommer:1993ce}.
In a physical volume $V\!\simeq\!(1.12\fm)^4\!\simeq\!1.57\fm^4$ there are
finite volume effects, but we expect them not to spoil the spectral analogy,
if it emerges at the given coupling.
Using the interpolation formula in~\cite{Guagnelli:1998ud} for the Wilson
gauge action, one finds that the pairs $(L\!=\!6, \be\!=\!5.66)$,
$(L\!=\!8, \be\!=\!5.79)$, $(L\!=\!12, \be\!=\!6.0)$,
$(L\!=\!16, \be\!=\!6.18)$ produce the desired volume (the first coupling
is slightly out of bound, the formula holds for $5.7\!\le\!\be\!\le\!6.57$).

\bigskip

Tables~\ref{tab:06060606}\,-\,\ref{tab:16161616} contain the absolute values
of the overlap indices on our $6^4, 8^4, 12^4, 16^4$ lattices.
As expected, the fraction of configurations on which the 1\,APE, 3\,APE, 1\,HYP
and 3\,HYP overlap operators agree on the charge increases with $\be$ (for
$\be\!\to\!\infty$, this fraction is one).
Table~\ref{tab:06060606} is interesting in another respect: The normal
``rule'' (at high $\be$) that additional smearing operations will reduce the
charge (here, we ignore the unsmeared column) is broken at low $\be$
(at $\be\!=\!5.66$ there are five such configurations).
Finally, Table~\ref{tab:08080808} also contains our test regarding the impact
of the projection parameter $\rh$ in (\ref{diracoverlap}).
On the first configuration, besides using $\rh\!=\!1$, we determined the
spectrum with $\rh\!=\!1.6$.
The standard and 1\,APE overlap operators prove sensitive to this shift, the
more vigorously filtered versions not.
This is fully compatible with the view that the charge ambiguity is an $O(a^2)$
effect; with strong UV-filtering the coefficient in front of the artefact gets
smaller.
Interestingly, $D^\mr{over}_\mr{63\,APE,\;63\,HYP}$ still yield reasonable
charge determinations, in spite of these operators being entirely non-local on
our small lattices.

\begin{table}
\begin{center}
\begin{tabular}{|l|cccccccc|}
\hline
$\be\!=\!5.66$&1\,APE&3\,APE&7\,APE&63\,APE&1\,HYP&3\,HYP&7\,HYP&63\,HYP\\
\hline
000 & 1 & 2 & 1 & 1 & 2 & 1 & 1 & 1\\
001 & 1 & 1 & 1 & 0 & 2 & 1 & 1 & 0\\
002 & 1 & 1 & 2 & 1 & 2 & 2 & 2 & 1\\
003 & 0 & 0 & 0 & 1 & 0 & 0 & 0 & 0\\
004 & 0 & 1 & 1 & 2 & 1 & 2 & 2 & 2\\
005 & 1 & 1 & 0 & 0 & 1 & 0 & 0 & 0\\
006 & 1 & 2 & 2 & 2 & 2 & 2 & 2 & 2\\
007 & 1 & 1 & 1 & 1 & 1 & 1 & 1 & 1\\
008 & 0 & 0 & 0 & 0 & 0 & 0 & 0 & 0\\
009 & 0 & 0 & 0 & 0 & 0 & 0 & 0 & 0\\
\hline
\end{tabular}
\end{center}
\vspace{-6mm}
\caption{\sl Number of zero modes of various overlap operators on our quenched
$6^4$ configurations. The 1\,APE, 3\,APE, 1\,HYP and 3\,HYP operators agree for
4 of the 10 configurations.}
\label{tab:06060606}
\end{table}

\begin{table}
\begin{center}
\begin{tabular}{|l|ccccccc|}
\hline
$\be\!=\!5.79$&orig&1\,APE&3\,APE&7\,APE&1\,HYP&3\,HYP&7\,HYP\\
\hline
000& 1(2) & 2(3) & 2(2) & 2(2) & 2(2) & 2(2) & 2(2)\\
001& 1 & 1 & 1 & 1 & 1 & 1 & 1\\
002& 0 & 1 & 0 & 1 & 1 & 1 & 1\\
003& 1 & 2 & 2 & 2 & 2 & 2 & 2\\
004& 0 & 1 & 1 & 1 & 1 & 1 & 1\\
005& 0 & 0 & 0 & 0 & 0 & 0 & 0\\
\hline
\end{tabular}
\end{center}
\vspace{-6mm}
\caption{\sl Number of zero modes of various overlap operators on our quenched
$8^4$ configurations. On the first configuration, the values for $\rh\!=\!1.6$
are in brackets. The 1\,APE, 3\,APE, 1\,HYP and 3\,HYP operators agree for 5 of
the 6 configurations.}
\label{tab:08080808}
\end{table}

\begin{table}
\begin{center}
\begin{tabular}{|l|ccccccc|}
\hline
\,$\be\!=\!6.0$\,&orig&1\,APE&3\,APE&7\,APE&1\,HYP&3\,HYP&7\,HYP\\
\hline
000& 1 & 1 & 1 & 1 & 1 & 1 & 1\\
001& 0 & 0 & 0 & 0 & 0 & 0 & 0\\
002& 1 & 1 & 1 & 1 & 1 & 1 & 1\\
003& 0 & 0 & 0 & 0 & 0 & 0 & 0\\
004& 1 & 1 & 1 & 1 & 1 & 1 & 1\\
005& 1 & 1 & 1 & 1 & 1 & 1 & 1\\
\hline
\end{tabular}
\end{center}
\vspace{-6mm}
\caption{\sl Number of zero modes of various overlap operators on our quenched
$12^4$ configurations. The 1\,APE, 3\,APE, 1\,HYP and 3\,HYP operators agree
for all 6 configurations.}
\label{tab:12121212}
\end{table}

\begin{table}
\begin{center}
\begin{tabular}{|l|ccccccc|}
\hline
$\be\!=\!6.18$&orig&1\,APE&3\,APE&7\,APE&1\,HYP&3\,HYP&7\,HYP\\
\hline
000 & 0 & 0 & 0 & 0 & 0 & 0 & 0\\
002 & 0 &---&---&---& 0 & 0 & 0\\
004 & 0 & 0 & 0 & 0 & 0 & 0 & 0\\
006 & 0 &---&---&---& 0 & 0 & 0\\
008 & 1 & 1 & 1 & 1 & 1 & 1 & 1\\
010 & 1 &---&---&---& 2 & 2 & 1\\
012 & 1 & 1 & 1 & 1 & 1 & 1 & 1\\
014 & 1 &---&---&---& 1 & 1 & 1\\
016 & 1 &---&---&---& 1 & 1 & 1\\
\hline
\end{tabular}
\end{center}
\vspace{-6mm}
\caption{\sl Number of zero modes of various overlap operators on our quenched
$16^4$ configurations. The 1\,APE, 3\,APE, 1\,HYP and 3\,HYP operators agree
for all 4 configurations on which they have all been determined.}
\label{tab:16161616}
\end{table}

We tested on the $8^4$ configurations whether the $SU(3)$-projection in the
APE~\cite{ape} or HYP~\cite{hyp} smearing recipes could be skipped.
The answer is negative:
With one smearing step, the impact is not so dramatic, but after two more
the staggered spectra are dramatically squeezed (rather than stretched out),
so that there is no agreement with the (filtered and projected) overlap
spectrum at all.

\begin{figure}
\vspace{-4mm}
\begin{center}
\epsfig{file=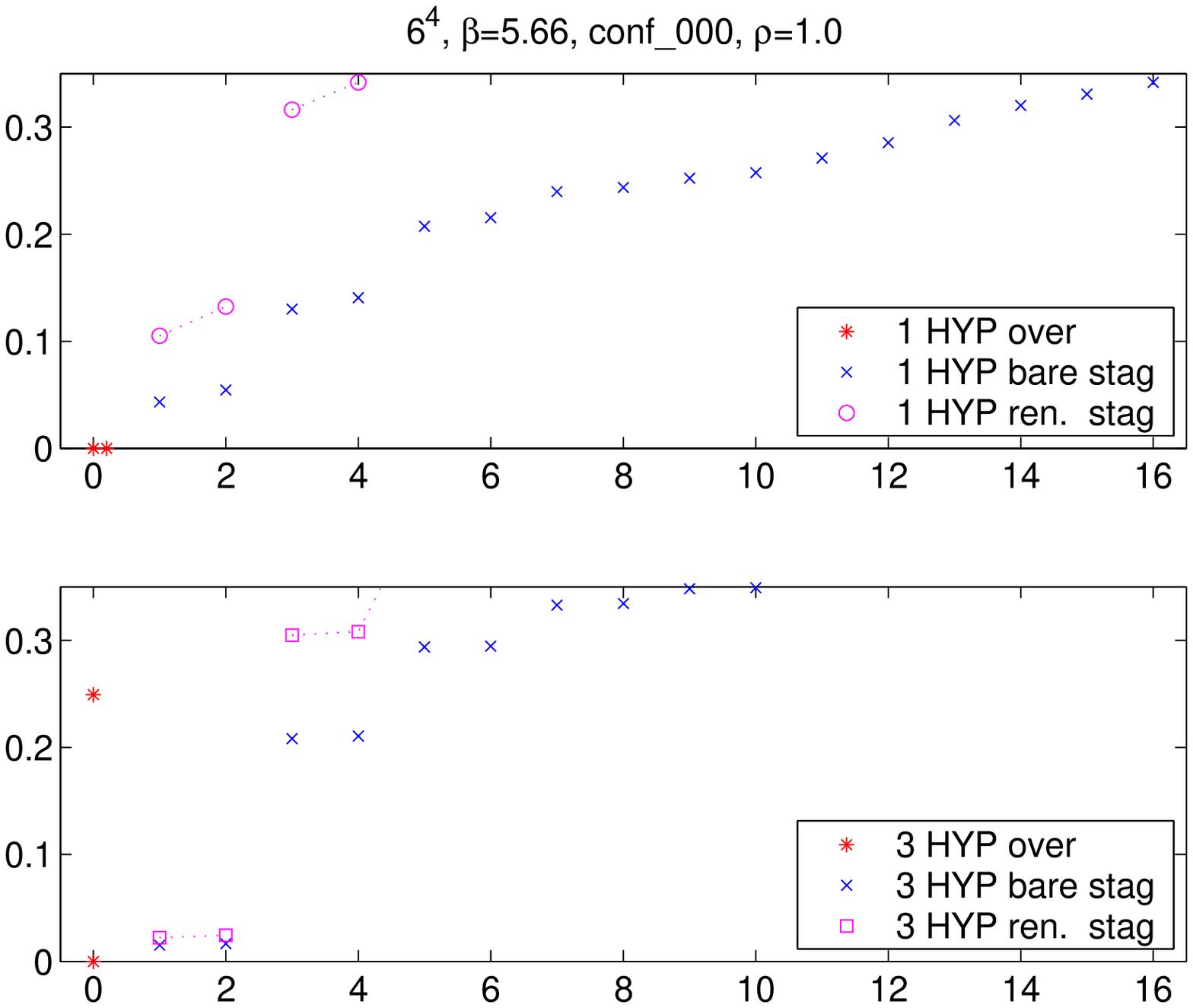,height=12.2cm}
\epsfig{file=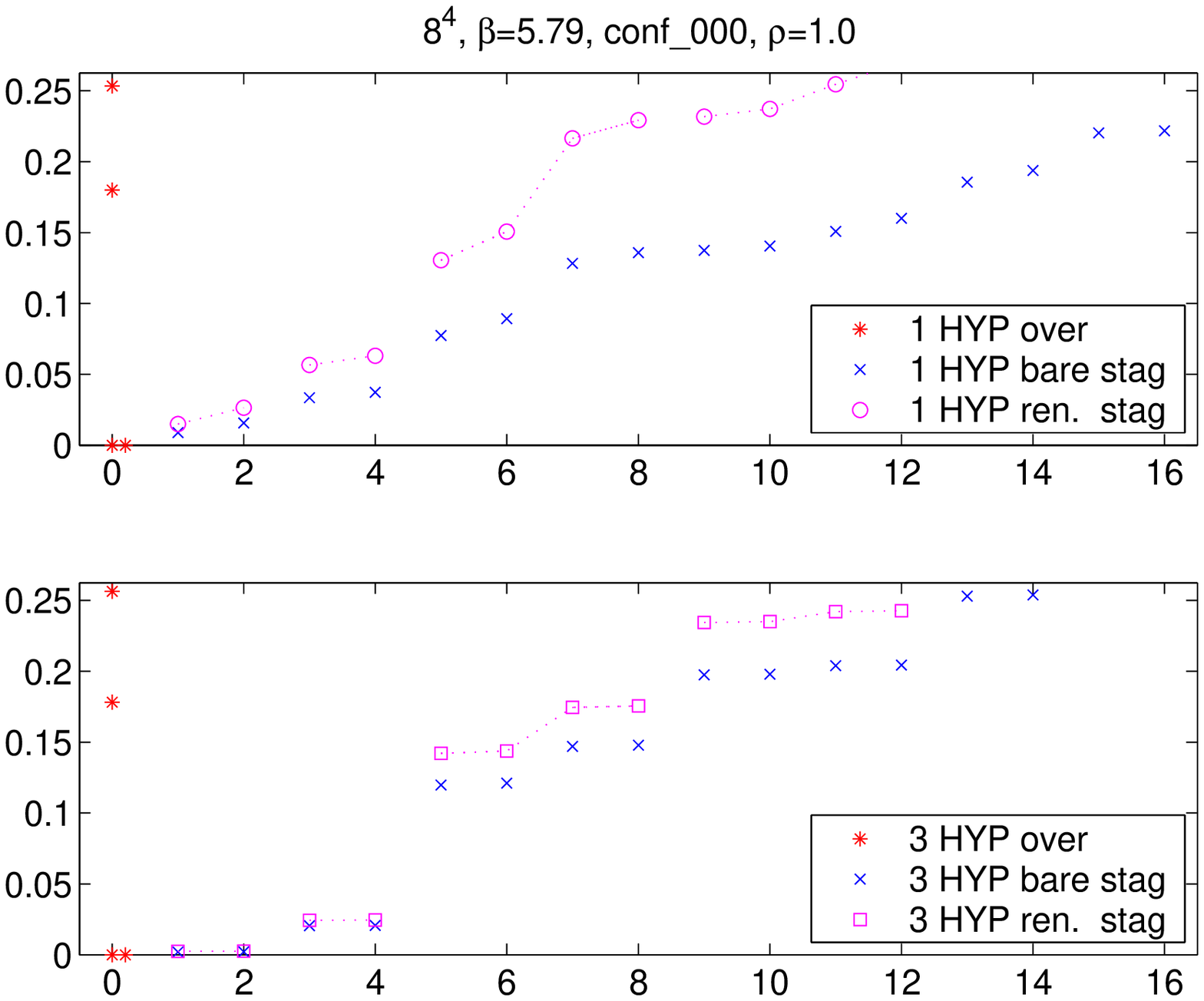,height=12.2cm}
\end{center}
\vspace{-9.5mm}
\caption{Staggered eigenvalues with/without relative normalization factor on
the first one of our $6^4$ and $8^4$ lattices, compared to the corresponding
overlap spectrum.}
\label{fig:0608_oneone}
\end{figure}

\begin{figure}
\vspace{-4mm}
\begin{center}
\epsfig{file=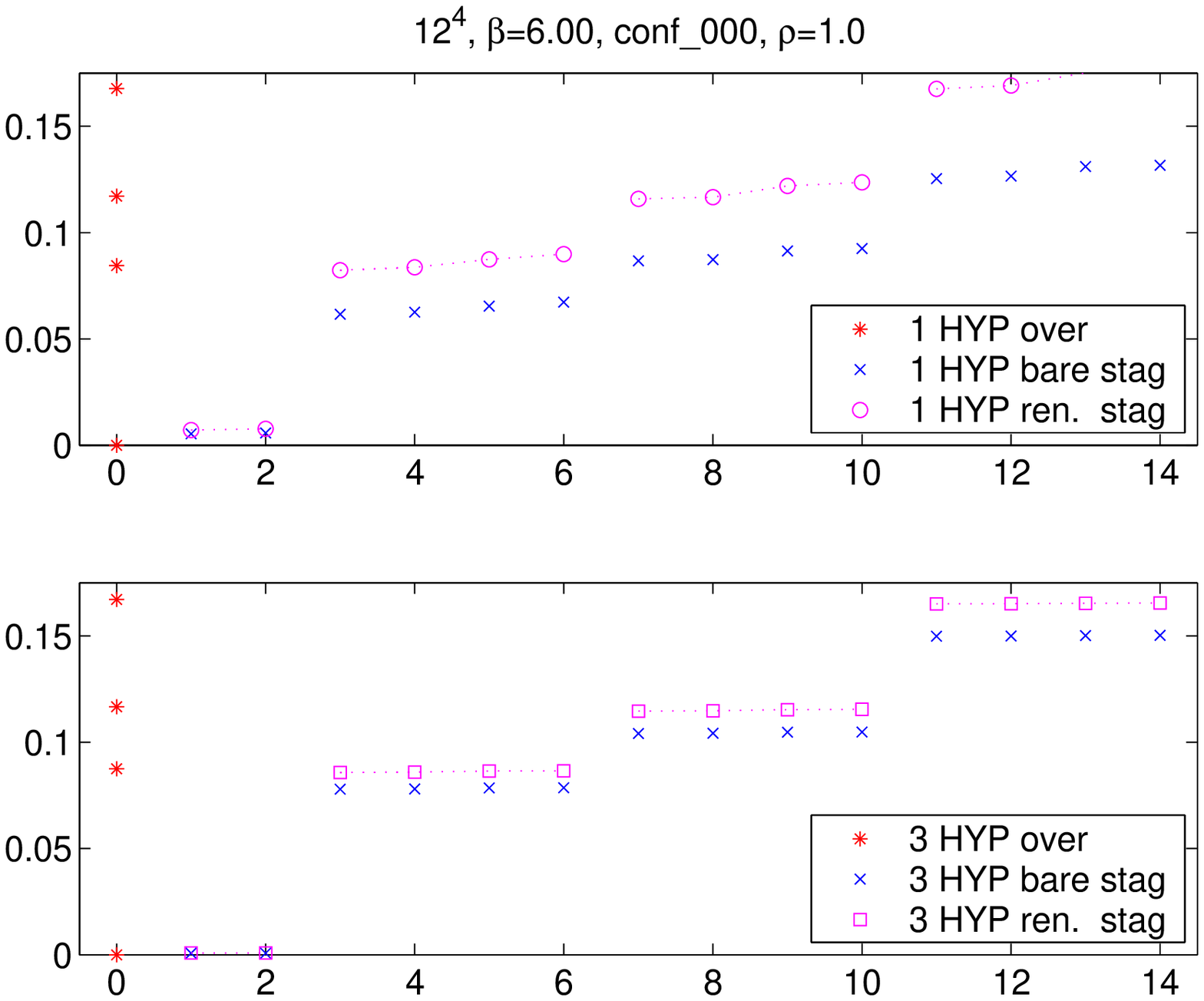,height=12.2cm}
\epsfig{file=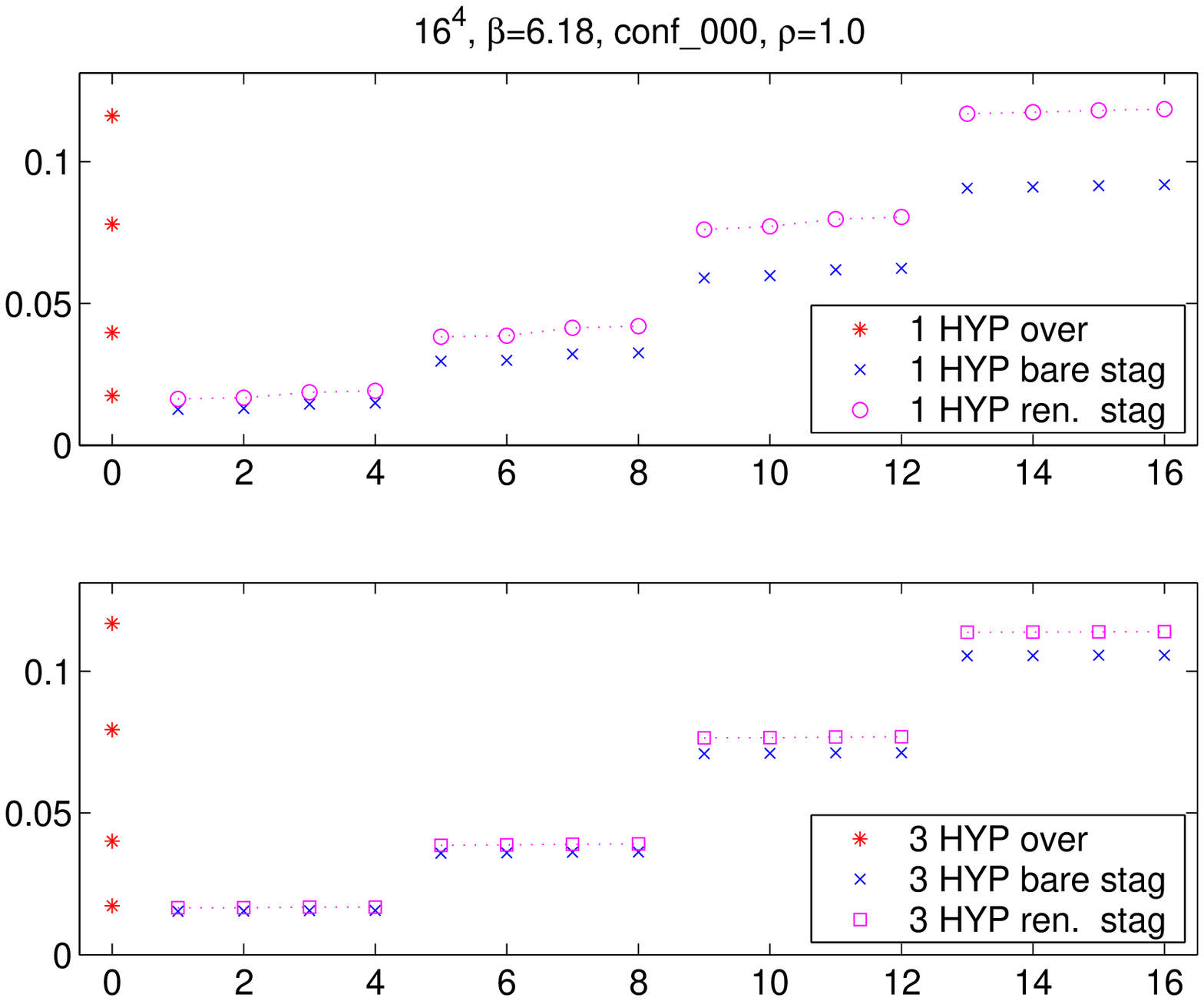,height=12.2cm}
\end{center}
\vspace{-9.5mm}
\caption{Staggered eigenvalues with/without relative normalization factor on
the first one of our $12^4, 16^4$ lattices, compared to the corresponding
overlap spectrum.}
\label{fig:1216_oneone}
\end{figure}

Contrary to the CU\_0001 case, we refrain from showing all spectra.
For the lack of a good criterion, we simply select the first configuration of
our $6^4$, $8^4$, $12^4$, $16^4$ chains, respectively.
In Figures~\ref{fig:0608_oneone} and \ref{fig:1216_oneone} the (ordered)
staggered eigenvalues (with projection) are plotted against their serial
number.
One clearly observes how the expected 4-fold near-degeneracy gets more
pronounced as $\be$ increases; at $\be\!=\!6.18$ there is a clear
``terrace~dynamics''.
This picture holds for $D^\mr{stag}_\mr{1\,HYP}$, and for
$D^\mr{stag}_\mr{3\,HYP}$ it is even more convincing.
On the left, the IR-spectrum of the overlap operator with the same type of
filtering is shown.
Finally, the appropriate relative normalization factor (slope in
Figs.~\ref{fig:0608_stagover} and \ref{fig:1216_stagover}, see below) is
applied to the staggered spectrum to make it a fair comparison.
The bottom line is that, at sufficiently weak coupling, it is appropriate
to identify the \emph{geometric mean of a near-degenerate quadruple\/} with a
single \emph{eigenvalue\/} of the overlap operator.
Whether there is, in addition, a relationship between the modes --~ideally,
one would like to give a construction which, in analogy to the one by Kogut and
Susskind, manages to (approximately)
\emph{thin out the degrees of freedom\/}~-- is a challenge for the future.

\bigskip

Our Figures~\ref{fig:0608_oneone} and \ref{fig:1216_oneone} cover a fixed
energy range up to $\sim\!370\MeV$.
They demonstrate that both the separation between would-be zero modes and
non-zero modes and the grouping into quadruples crucially depend on the lattice
spacing~--~below $a^{-1}\!\simeq\!2\GeV$ even a high level of filtering does
not help.
And, by comparing to the filtered overlap we see a
\emph{quantitative~agreement\/} for $a^{-1}\!>\!2-2.5\GeV$.


\subsection{Correlation between staggered and overlap eigenvalues}

To check whether the quartic-rooted filtered staggered operator really
reproduces the spectrum of the filtered overlap one may plot them, eigenvalue
by eigenvalue, against each other.

\begin{figure}
\vspace{-4mm}
\begin{center}
\epsfig{file=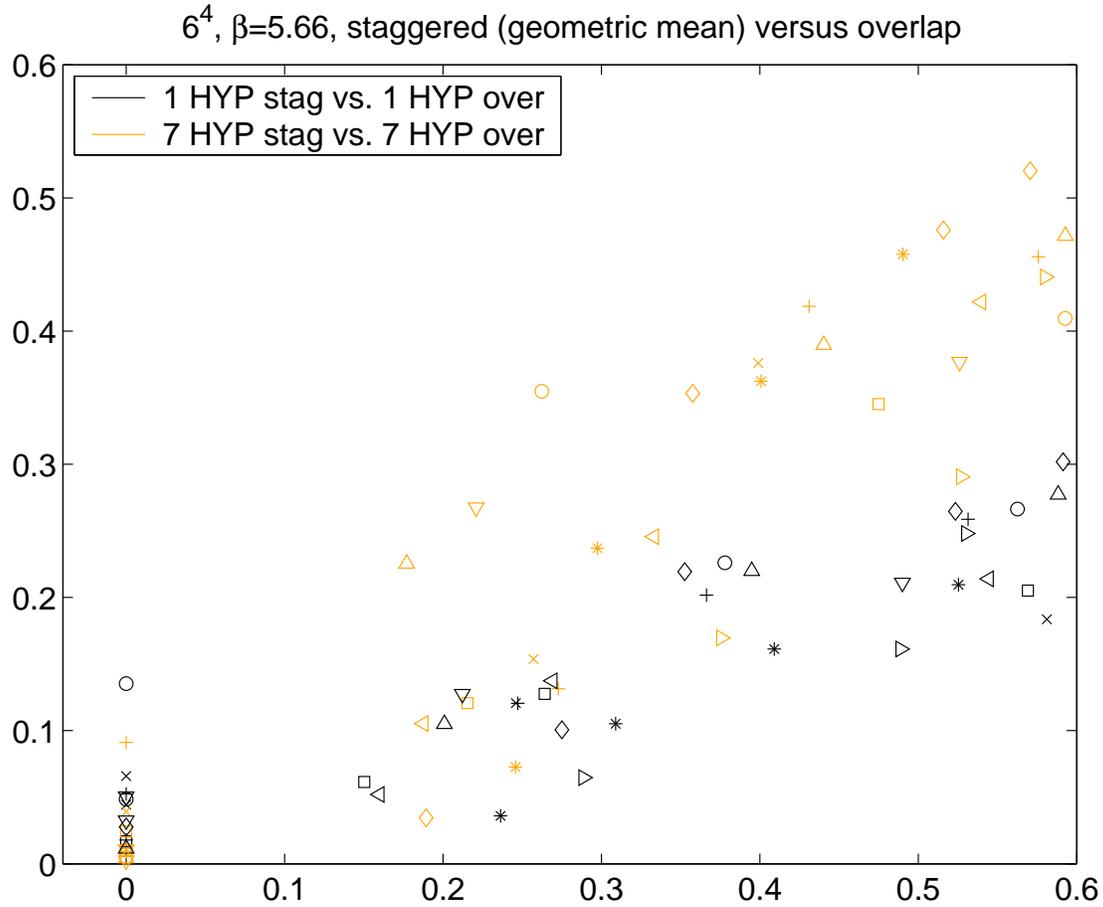,height=12.2cm}
\epsfig{file=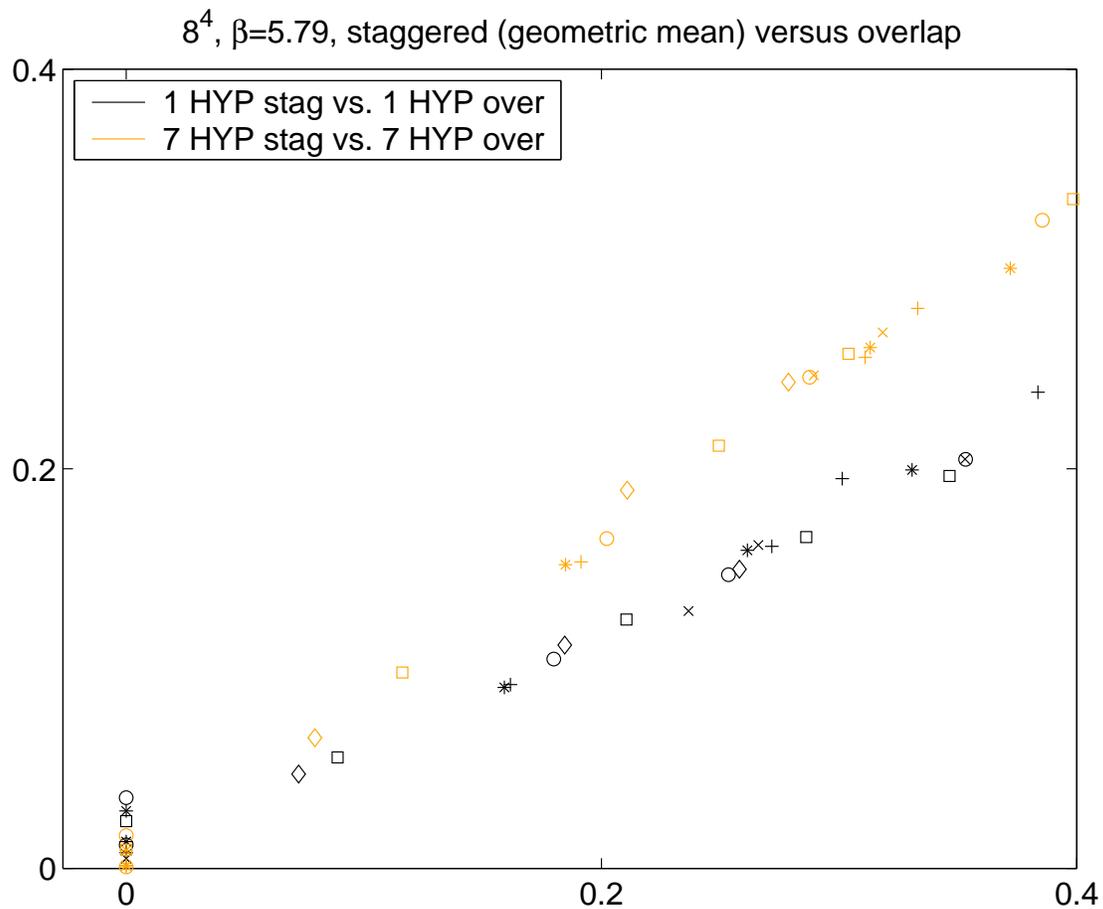,height=12.2cm}
\end{center}
\vspace{-9.5mm}
\caption{Quartic-rooted staggered versus overlap modes at $\be\!=\!5.66$ and
$\be\!=\!5.79$. Different symbols refer to different configurations.}
\label{fig:0608_stagover}
\end{figure}

\begin{figure}
\vspace{-4mm}
\begin{center}
\epsfig{file=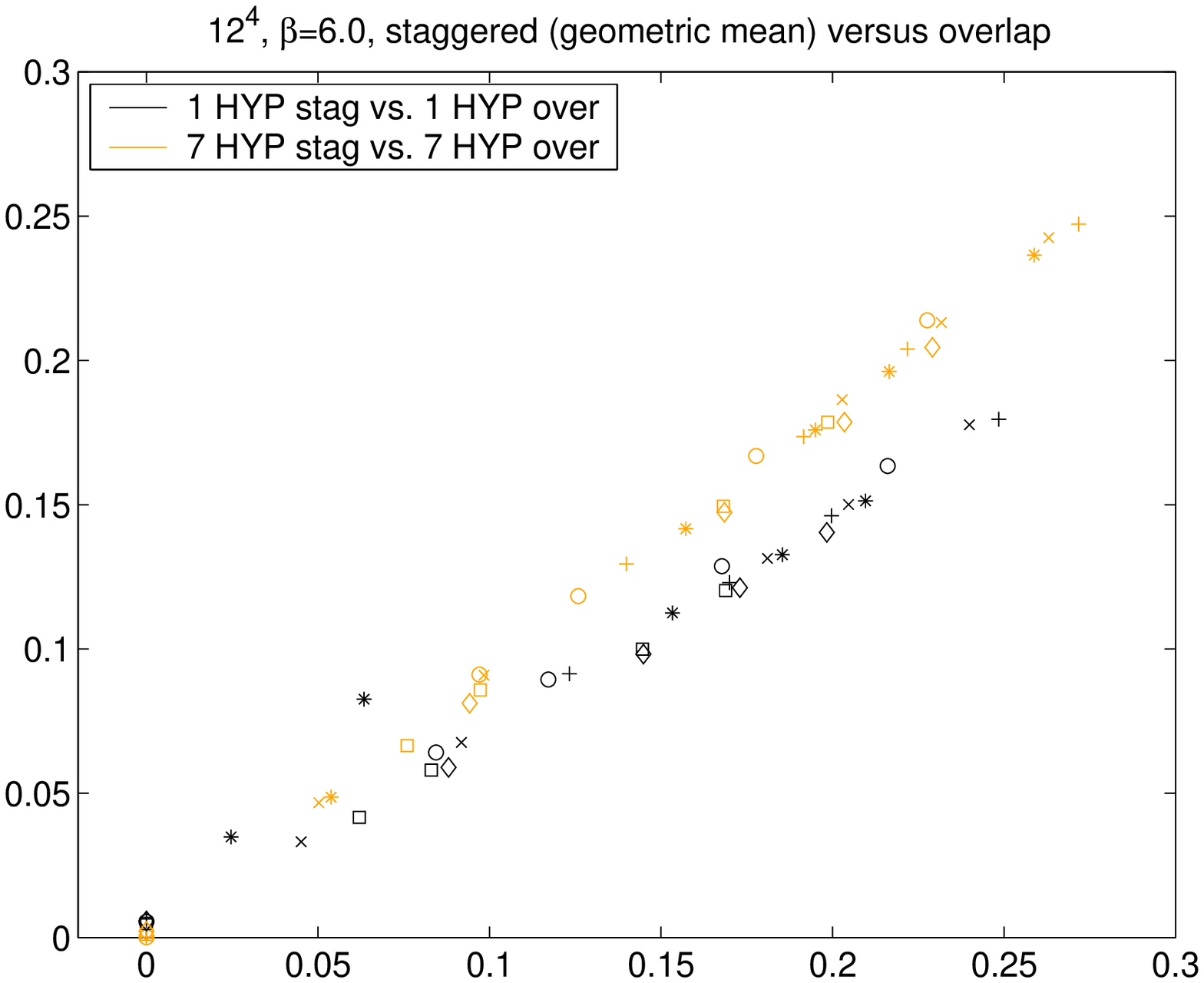,height=12.2cm}
\epsfig{file=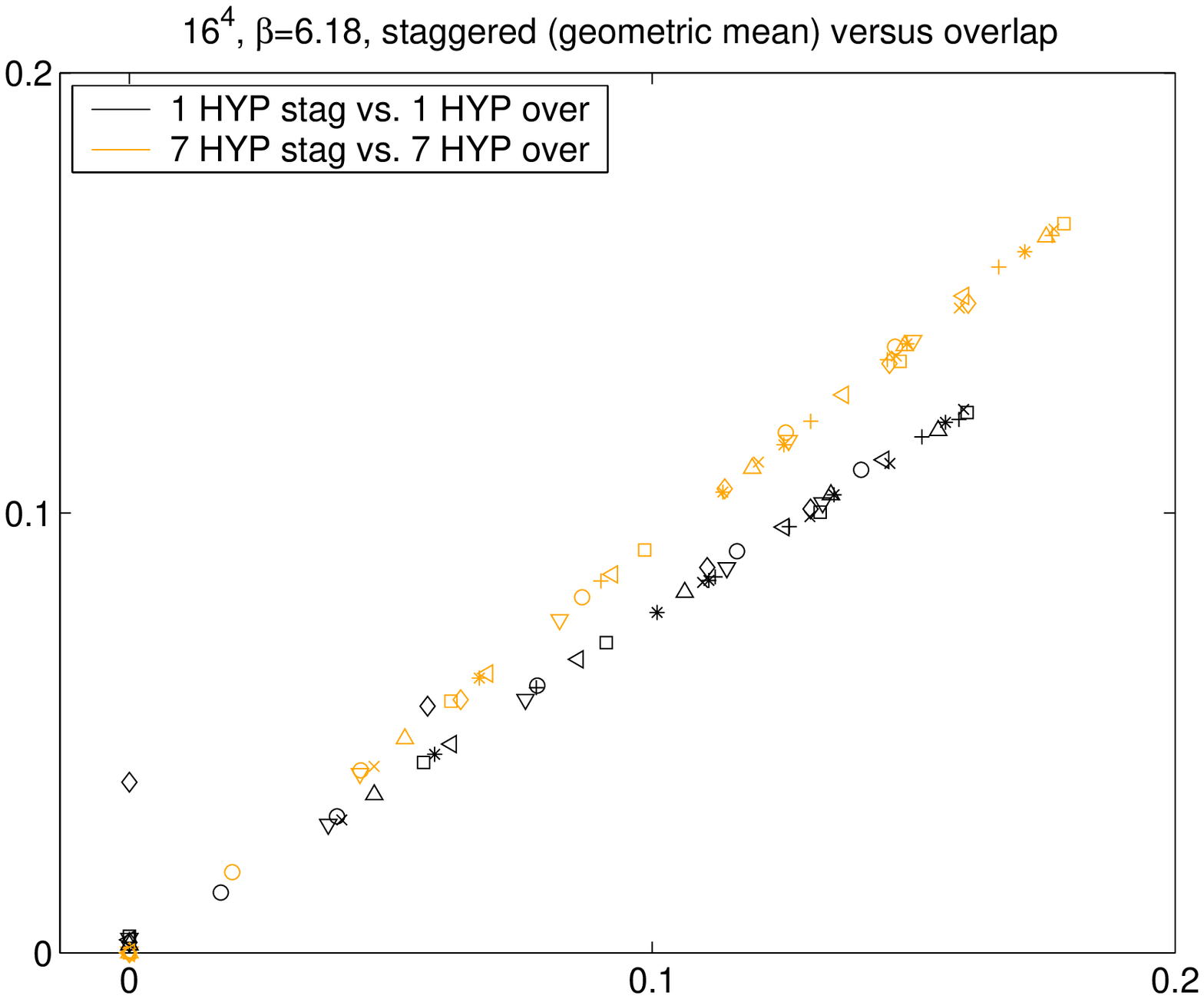,height=12.2cm}
\end{center}
\vspace{-9.5mm}
\caption{Quartic-rooted staggered versus overlap modes at $\be\!=\!6.0$ and
$\be\!=\!6.18$. Different symbols refer to different configurations.}
\label{fig:1216_stagover}
\end{figure}

Restricting ourselves to the positive half of the spectrum, we group the
staggered eigenvalues into two classes:
The would-be zero modes (we name them $0\!<\!\ze_1\!<\!\ze_2\!<\!...$)
and the (true) non-zero modes (denoted by $\la_1\!<\!\la_2\!<\!...$ with
$\ze_{\mr{max}=2|q|}\!<\!\la_1$).
For the $n$-th would-be zero mode we take $\sqrt{\ze_{2n-1}\ze_{2n}}$, since
$(-\ze_2, -\ze_1, \ze_1, \ze_2)$ is the substitute of the first zero mode.
The $m$-th non-zero mode is associated with
$(\la_{4m-3}\cdot...\cdot\la_{4m})^{1/4}$.
This classification is a strict function of $q$ determined via the overlap
against which one compares.

These geometric means are then plotted against the overlap eigenvalues
$\hat\la$, results being shown in
Figs.~\ref{fig:0608_stagover}-\ref{fig:1216_stagover}.
Here, every symbol corresponds to an individual configuration.

What one sees immediately, is that the agreement gets much better with
increasing $\be$.
In three graphs there is a ``pile'' of would-be zero modes above the origin.
These are classified so, because the overlap finds a non-zero index, but then
they happen to be not-so-small.
Since the overlap operator itself has an $O(a^2)$ ambiguity%
\footnote{With another $\rh$ or an alternative smearing level a different
number of staggered modes might have been identified as would-be zero modes.
\label{foot:thisone}},
this phenomenon is expected to become rare at weak coupling.
Unfortunately, even at $\be\!=\!6.18$ we have one configuration with such a
charge ambiguity.
The pile also shows up when one overlap is plotted versus
another one, say $D^\mr{over}_\mr{1\,HYP}$ vs. $D^\mr{over}_\mr{7\,HYP}$.
On the other hand, the fact that one configuration in the $\be\!=\!6.0$
ensemble goes astray is due to UV-noise which does not affect the charge.
One learns that at this coupling even the 1\,HYP operator suffers from
large UV-fluctuations in its IR-spectrum.

Switching in Figs.~\ref{fig:0608_stagover},~\ref{fig:1216_stagover} from the
1\,HYP level (black) to 7\,HYP staggered vs.\ 7\,HYP overlap (light), the
correlation gets tighter and the relative normalization factor $Z$ moves closer
to 1.
Increasing, on the staggered side, the smearing level to 7\,HYP, but staying,
on the overlap side, at 1\,HYP (i.e.\ $D^\mr{stag}_\mr{7 HYP}$ vs.\
$D^\mr{over}_\mr{1 HYP}$), the $Z$-factor still moves towards 1, but the
correlation does \emph{not~improve\/} (not shown).
We plan to check that the tight correlation between the lowest eigenmodes
extends to the whole determinant, i.e. to verify
eqns.~(\ref{fundamental1}, \ref{fundamental2}) directly.

Once $\be$ is so large that the pile of not-so-small would-be zero modes
is gone and a clear gap to the non-zero modes is observed
(cf.\ Fig.~\ref{fig:1216_stagover}), the filtered staggered fermions will
satisfy an \emph{approximate index theorem\/}.
This has been demonstrated in great detail in 2D~\cite{Durr:2003xs}, and
Refs.~\cite{Damgaard:1999bq,Follana:2004sz} have added evidence that this holds
in 4D, too.

Last but not least, the mode-by-mode correspondence implies that (filtered)
staggered fermions on sufficiently fine lattices reach agreement with the
prediction of random matrix theory, as it has been observed for overlap
fermions \cite{Giusti:2003gf}.


\subsection{Scaling analysis for eigenvalue-splitting pseudo-observables}

To define the fuzziness of the lowest would-be zero mode (if present) we form
\begin{eqnarray}
\Delta_0&=&{\sqrt{\ze_2\ze_1} \ovr
(\la_4\la_3\la_2\la_1)^{1/4}-\sqrt{\ze_2\ze_1}}
\;.
\label{defzero}
\end{eqnarray}
%
%
To define the fuzziness of the first (true) non-zero mode we construct
\begin{eqnarray}
\Delta_1&=&{\sqrt{\la_4\la_3}-\sqrt{\la_2\la_1} \ovr
(\la_8\la_7\la_6\la_5)^{1/4}-(\la_4\la_3\la_2\la_1)^{1/4}}
\;.
\label{defposi}
\end{eqnarray}
These expressions quantify the non-degeneracy of a quadruple on the scale of
its distance to the next one, and tend to zero if the four-fold degeneracy
becomes exact.

\begin{figure}
\vspace{-2mm}
\begin{center}
\epsfig{file=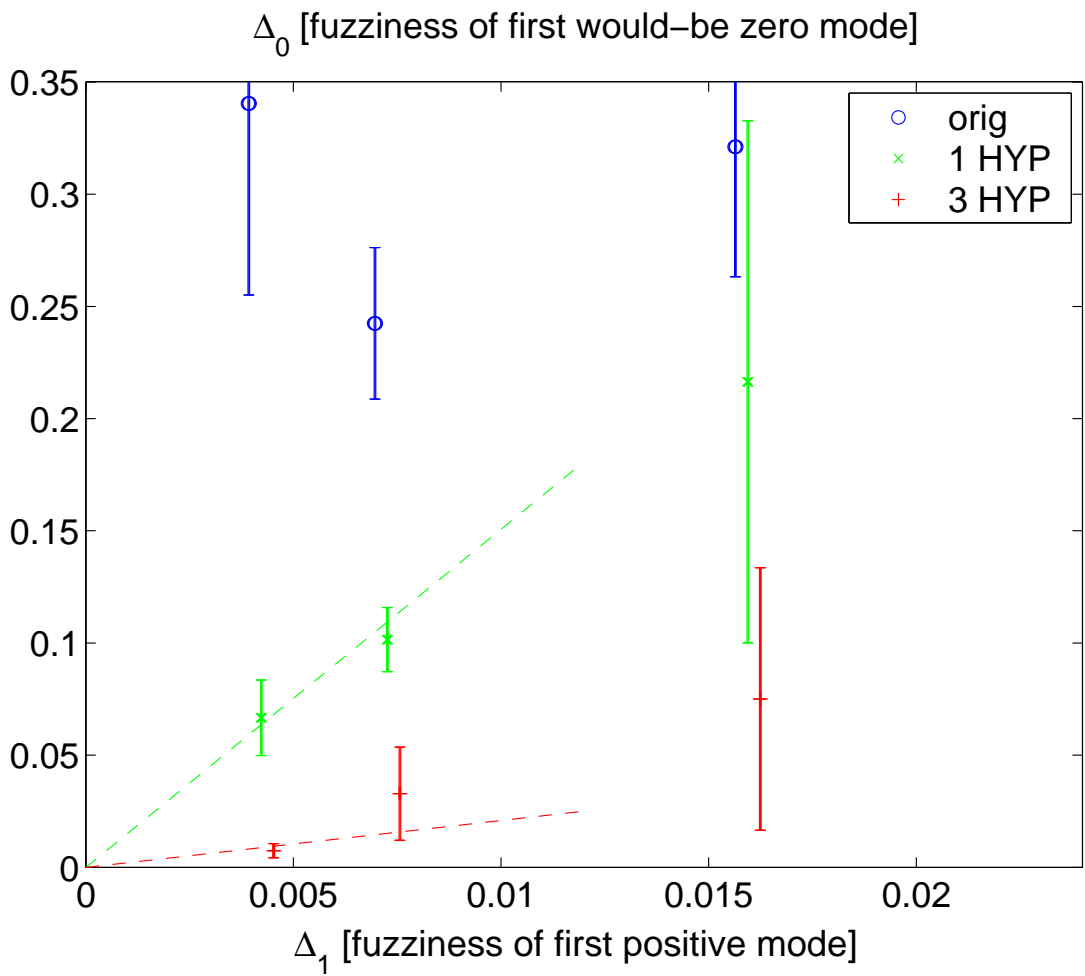,width=14.4cm}\\
\epsfig{file=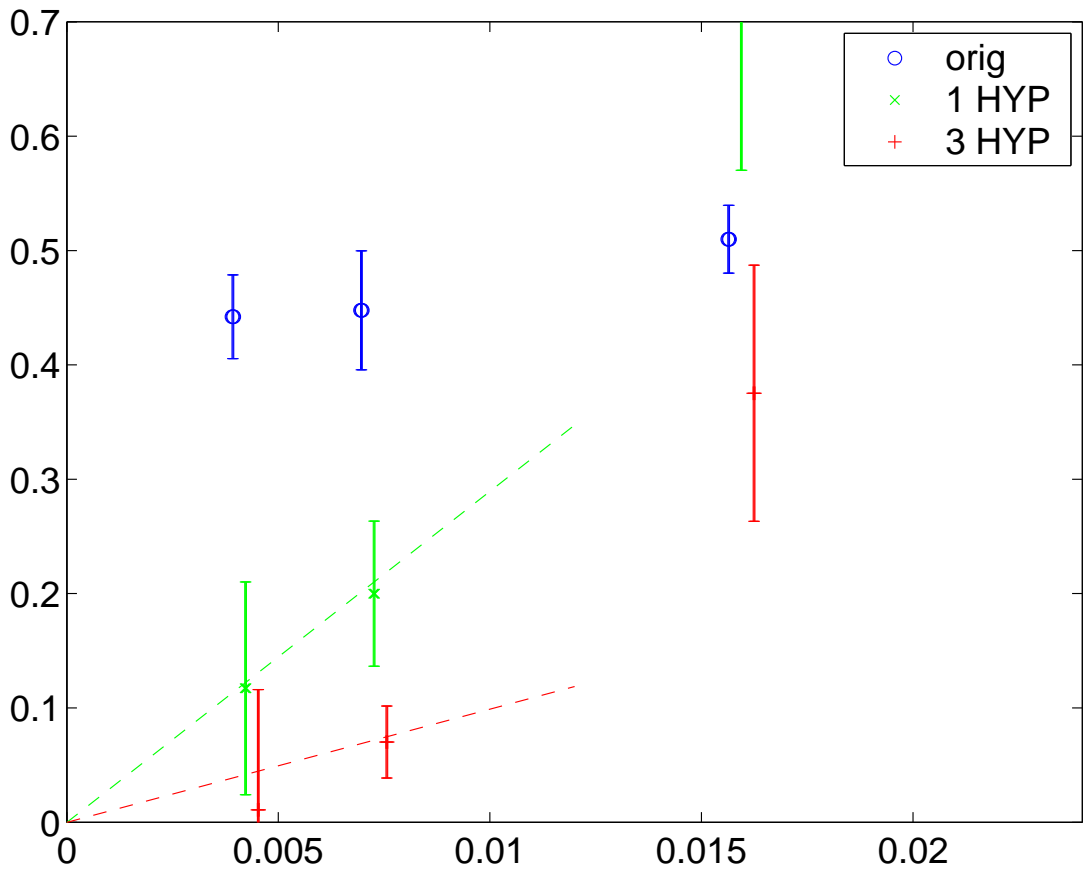,width=14.4cm}
\end{center}
\vspace{-9mm}
\caption{Pseudo-observables $\Delta_0$ and $\Delta_1$ vs.\ $1/L^2$ for the
matched $8^4, 12^4, 16^4$ lattices.}
\label{fig:summary}
\end{figure}

If the non-degeneracy of the staggered eigenmodes and the fuzziness of
the would-be zero modes are just $O(a^2)$ artefacts (though, numerically, they
may be so dramatic that without UV-filtering not even the faintest analogy is
seen at accessible couplings), one would expect that the pseudo-observables
(\ref{defposi}) and (\ref{defzero}) vanish (asymptotically) in proportion to
$a^2$.

Fig.~\ref{fig:summary} contains our results for the $8^4$, $12^4$, and $16^4$
lattices (the points for the $6^4$  geometry were omitted since they are
definitely not in the scaling regime).
Given the smallness of our samples, we have opted for the median instead of
the arithmetic mean -- this estimator is biased, but far more robust against
outliers.
Indeed, the 1\,HYP and 3\,HYP data seem to be consistent with the hypothesis
of asymptotic scaling -- although our data do, of course, not pin down the
exponent.
A linear fit (versus $a^2$) through the $L=16, 12$ points (constrained to go
through 0, i.e.\ with 1 d.o.f.) yields an acceptable
$\chi^2$, and its extension to the right seems compatible with the $L=8$ data.
On the other hand, the unfiltered data are essentially flat, and there is
no way to fit them in this manner.
This is precisely the scenario mentioned in the introduction.
While we have no doubts that --~moving to much weaker couplings~-- even the
unsmeared staggered operator's pseudo-observables $\Delta_{0,1}$ would
eventually turn down and reach the origin linearly, the big difference with
the 1\,HYP and 3\,HYP versions is that this happens at
\emph{accessible~couplings\/}.

Furthermore, the lesson from 2D is that a typical $\sqrt{\ze_{2q}\ze_{2q-1}}$
of the massless operator gives a reliable estimate down to which quark mass the
massive staggered Dirac operator yields trustworthy results~\cite{Durr:2003xs}.


\subsection{Technical bonus}

It is interesting to note that the construction of the UV-filtered overlap
operator is much cheaper than the standard version.
For the unsmeared CU configurations the condition number of 
$H^2=(\gaf D_{-1}^\mr{W})^2$ was in the range  $10^4\!-\!10^8$.
After projecting out the 11 lowest eigenmodes, this figure gets reduced to
$\sim\!350-450$, and the degree of the associate Chebychev polynomial is
$\sim\!400$.
After 3\,HYP steps, the condition number is $\sim\!140\!-\!15'000$ and
$\sim\!60\!-\!100$, before and after projection, respectively, and the required
degree of the Chebychev polynomial on the subspace orthogonal to the 11 lowest
modes is $\sim\!60$.

Note that this impact on the degree of the polynomial is seen after the
lowest $11$ eigenvectors have been projected out.
This strengthens the impression that one gets from comparing Fig.\,2
in Ref.~\cite{Hasenfratz:2002rp} to Fig.\,3 in Ref.~\cite{Jansen:2003jq}.
In the latter case, the standard (unfiltered) Wilson kernel is evaluated on
pure gauge backgrounds generated with the Iwasaki/DBW2 action, and the lowest
eigenvalues of $H^2$ get considerably lifted compared to the Wilson gauge
action case.
However, after a fixed number of eigenmodes have been projected out, this
advantage is considerably reduced (see Fig.~3 of Ref.~\cite{Jansen:2003jq}),
since the bulk part of the spectrum is less sensitive to the improvement
coefficient than the bottom end of the eigenvalue distribution.
In Ref.~\cite{Hasenfratz:2002rp} the fermion action is changed, and this has
a huge impact even after a fixed number of eigenvectors is projected out.
In line with this, we observe that the \emph{bulk\/} of the eigenmodes of $H^2$
for a HYP smeared gauge configuration (which we see as a filtered fermion on
the original background) starts at significantly higher values.
This holds, regardless whether this configuration was produced with the Wilson
gauge action or stems from a full QCD ensemble.
Therefore, the construction of the overlap operator with a UV-filtered kernel
is considerably cheaper than for a standard Wilson kernel even after projecting
out the few lowest eigenmodes of $H^2$.


\section{Summary and Outlook}


We have studied a class of UV-filtered staggered and overlap operators that are
built with standard techniques, comparing the IR-part of their spectra with
and without filtering.
We close with a summary of our findings and a few comments on their
implications.

\begin{enumerate}

\item
With sufficient filtering and on fine enough lattices, the staggered fermions
develop an ability to separate would-be zero modes from non-zero modes, and
arrange them in near-degenerate groups of 4.
The second point is that both the number of zero-modes and the quadruples
reproduce, up to $O(a^2)$ effects, the findings of the filtered overlap
operator. This implies an approximate index theorem for filtered staggered
quarks.

\item
For both aspects, the ``critical'' coupling is (in quenched Wilson terms)
around $\be_\mr{W}\!=\!6.0$.
Below, the agreement vanishes rapidly, and there is no way to compensate for
this by resorting to higher smearing levels.

\item
The agreement, at the level of a single configuration, with the low-energy
spectrum of the overlap operator implies that UV-filtered staggered fermions
will eventually reproduce the correct random-matrix-theory universality class
and the correct continuum limit of topological susceptibility and associate
flavor-singlet observables, as a corollary.

\item
A rudimentary scaling analysis of two pseudo-observables designed to capture
the fuzziness of the staggered would-be zero modes and non-zero modes,
respectively, suggests that our data are consistent with the hypothesis that
these staggered artefacts vanish like~$O(a^2)$.
This, together with the quantitative agreement with the corresponding overlap 
eigenvalues, forms evidence in favor of using the fractional power
(\ref{fundamental2}) of the filtered staggered determinant.

\item
An observation important for quenched overlap spectroscopy is that the filtered
overlap is much cheaper than the unfiltered one.
This is a distinctive feature of the kernel in (\ref{diracoverlap}), and not
tied to any particular action used to generate the background.

\item
Our line-up and correlation plots at weak coupling
(Figs.~\ref{fig:1216_oneone} and \ref{fig:1216_stagover}) bear the promise to
trigger a new development in lattice QCD.
With such an excellent one-to-four agreement of the low-lying spectra, it might
be possible to run a dynamical $\Nf\!=\!2$ staggered simulation, performing the
final accept/reject-step with the overlap action (or, alternatively, to
reweight the ensemble to $\Nf\!=\!2$ overlap flavors).
In other words, the perspective is to perform a fully fledged dynamical
overlap or domain-wall study essentially at the costs of a staggered run.

\end{enumerate}


\subsection*{Acknowledgments}

S.D.\ and U.W.\ would like to acknowledge stimulating discussions with Karl
Jansen, Ch.H. with Laurent Lellouch and Leonardo Giusti.
S.D.\ is supported by DFG in SFB/TR-9,
Ch.H.\ is supported by EU grant HPMF-CT-2001-01468.



\end{document}